\documentclass[prb,twocolumn,superscriptaddress,showpacs,unsortedaddress]{revtex4}
\usepackage{graphicx}
\usepackage{dcolumn}
\usepackage{bm}
\usepackage{subfigure}
\usepackage{color}
\usepackage{amsmath}
\usepackage{amssymb}
\usepackage{multirow}

\begin{document}
\title{Electric field control of multiferroic domains in Ni$_3$V$_2$O$_8$ imaged by X-ray polarization
enhanced topography}

\author{F. Fabrizi}
\affiliation{European Synchrotron Radiation Facility, Bo\^{i}te Postale 220, 38043 Grenoble, France }
\affiliation{London Centre for Nanotechnology, University College London, 17-19 Gordon Street, London WC1H 0AH, UK}
\author{H. C. Walker}
\affiliation{European Synchrotron Radiation Facility, Bo\^{i}te Postale 220, 38043 Grenoble, France }
\affiliation{London Centre for Nanotechnology, University College London, 17-19 Gordon Street, London WC1H 0AH, UK}
\author{L. Paolasini}
\affiliation{European Synchrotron Radiation Facility, Bo\^{i}te Postale 220, 38043 Grenoble, France }
\author{F. de Bergevin}
\affiliation{European Synchrotron Radiation Facility, Bo\^{i}te Postale 220, 38043 Grenoble, France }
\author{T. Fennell}
\affiliation{London Centre for Nanotechnology, University College London, 17-19 Gordon Street, London WC1H 0AH, UK}
\affiliation{Institut Laue-Langevin, 38042 Grenoble, France}
\author{N. Rogado}
\affiliation{Department of Chemistry and Princeton Materials Institute, Princeton University, Princeton, New Jersey 08544, USA}
\author{R.J. Cava}
\affiliation{Department of Chemistry and Princeton Materials Institute, Princeton University, Princeton, New Jersey 08544, USA}
\author{Th. Wolf}
\affiliation{Forschungszentrum Karlsruhe, Institut f\"ur Festk\"orperphysik, D-76021 Karlsruhe, Germany}
\author{M. Kenzelmann}
\affiliation{Laboratory for Developments and Methods, Paul Scherrer Institute, CH-5232 Villigen, Switzerland}
\author{D. F. McMorrow}
\affiliation{London Centre for Nanotechnology, University College London, 17-19 Gordon Street, London WC1H 0AH, UK}

\date{\today}

\begin{abstract}
The magnetic structure of multiferroic Ni$_3$V$_2$O$_8$ has been investigated using non-resonant X-ray magnetic scattering. Incident circularly polarized X-rays combined with full polarization analysis of the scattered beam is shown to yield high sensitivity to the components of the cycloidal magnetic order, including their relative phases. New information on the magnetic structure in the ferroelectric phase is obtained, where it is found that the magnetic moments on the ``cross-tie'' sites are quenched relative to those on the ``spine'' sites. This implies that the onset of ferroelectricity is associated mainly with spine site magnetic order. We also demonstrate that our technique enables the imaging of multiferroic domains through polarization enhanced topography. This approach is used to image the domains as the sample is cycled by an electric field through its hysteresis loop, revealing the gradual switching of domains without nucleation.
\end{abstract}

\pacs{75.25.-j, 75.85.+t, 77.80.Dj}

\maketitle
\section{Introduction}
X-ray imaging continues to find growing utility driven in part by the availability of intense beams from synchrotron and free-electron sources, and in part by a growing awareness of how to exploit various contrast mechanisms and imaging modes. Imaging techniques range from those that depend on monitoring the absorption, or phase shift \cite{David02,Weitkamp05,Pfeiffer08}, to diffraction of an x-ray beam, including the possibilities offered by coherent beams \cite{Miao99,Robinson01} and holography \cite{McNulty92,Eisebitt04}. Here we introduce a technique where imaging proceeds by measuring the polarization state of a magnetically scattered photon through determination of its Stokes parameters. We demonstrate that when applied to multiferroic Ni$_3$V$_2$O$_8$ it provides topographic images of the spatial distribution of magnetic domains as the ferroelectric domains are cycled through a hysteresis loop by an applied electric field. Our technique has potential for imaging domains in multiferroic devices and other classes of correlated electron systems which are characterised by unconventional  or coupled order parameters.

The discovery of strongly coupled magnetic and ferroelectric order parameters in TbMnO$_3$ \cite{KimuraTMO}, and subsequently in other families of compounds \cite{Hur,heyer}, has led to a renaissance of interest in magnetoelectric materials. It is now established that TbMnO$_3$ typifies a new class of  materials in which the onset of ferroelectricity is driven by the formation of non-collinear, cycloidal magnetic order which breaks spatial inversion symmetry. One of the outstanding challenges in this field is the development of methods capable of imaging multiferroic domains, and most especially their evolution under applied external fields. However, the imaging of ferroelectric and magnetic domains has largely remained two separate fields, with a few exceptions\cite{Fiebignat,Schierle}. Imaging of ferroelectric domains is well established, and can be achieved through a variety of probes including X-ray charge scattering \cite{Rejmankova,Gonzalez}, and atomic force microscopy \cite{Hong}. In contrast, the imaging of antiferromagnetic domains has emerged in more recent times, driven by the availability of highly brilliant X-ray beams from synchrotron sources. The high spatial resolution attainable with such sources has enabled new imaging methods based on either the absorption \cite{Nolting} or scattering \cite{Evans,Eisebitt04,Shpyrko} of an X-ray beam through various processes which yield sensitivity to the antiferromagnetic order.

\begin{figure*}
\centering
\includegraphics[width=.8\textwidth]{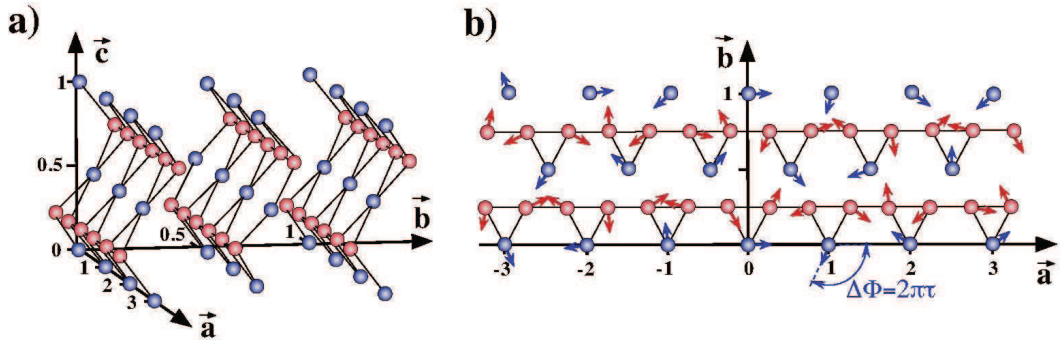}
\caption{Crystallographic and low temperature incommensurate (LTI) magnetic structure of the nickel sublattice in Ni$_3$V$_2$O$_8$. (a) Schematic of the two nickel sites: spine (red) and cross-tie (blue) (b) Projection onto the a-b plane of the proposed LTI magnetic structure of the spine sites $\mathbf{m}^{SP}=(1.6\pm0.1, 1.3\pm0.1, 0)\mu_\mathrm{B}$ and the cross-tie sites $\mathbf{m}^{CT}=(2.2\pm0.1, 1.4\pm0.1, 0)\mu_\mathrm{B}$, as deduced from neutron diffraction measurements \cite{kenzelmann06}.\label{fig:cryst}}
\end{figure*}

For the new class of multiferroics exemplified by TbMnO$_3$, the potential of X-rays for providing information on the multiferroic state has been highlighted by several studies \cite{Beutier,Ewings,Feyerherm,mannix,Strempfer,Wilkins09}. In particular, we have demonstrated earlier that the handedness of circularly polarised X-rays naturally couples to the handedness of cycloidal spin structures, and that the change in polarization of an X-ray photon undergoing a non-resonant magnetic  scattering process encodes detailed information on the cycloidal magnetic order\cite{Fabrizi}. The polarization state of a photon may be described by the Stokes parameters $\bf P$=$(P_1,P_2,P_3)$ \cite{Blume,deBergevin}, where a linear polarization analyser may be used to determine the Stokes parameters $P_1$ and $P_2$. By definition $P_1$ describes the component of the linear polarisation in and perpendicular to the scattering plane, $P_2$ the oblique linear polarisation and $P_3$ the circular polarisation. Here we demonstrate the imaging of magnetic domains, and hence the ferroelectric
domains, in multiferroic Ni$_3$V$_2$O$_8$ where the imaging contrast derives from the use of circularly polarized X-rays to excite non-resonant magnetic scattering processes, and topographic imaging is performed by mapping the spatial variation of the Stokes parameters.

Ni$_3$V$_2$O$_8$ (NVO) is an example of the new class of magnetoelectric multiferroics.  It is a magnetic insulator with a structure characterised  by planes of Ni$^{2+}$ $S=1$ spins arranged on a buckled kagom\'{e} staircase structure, resulting in two inequivalent Ni sites: ``cross-tie'' and ``spine''-type, as shown in Fig.\  \ref{fig:cryst}(a). Whilst the ideal kagom\'{e} lattice is the canonical example of a frustrated system, the deviation from this geometry in NVO leads to additional interactions relieving the frustration and producing a rich magnetic phase diagram \cite{Lawes04,Lawes05,kenzelmann06}.  Below $T_L=6.3$~K, in the low temperature incommensurate (LTI) phase, the spins on both site types are reported to be arranged in a cycloid in the $a-b$ plane with propagation vector ($\tau\approx0.27$ 0 0), breaking spatial inversion symmetry. It is in this phase that a spontaneous electric polarisation develops parallel to the $b$-axis \cite{Lawes05}.  The magnetic structure in the LTI phase derived from the neutron diffraction studies \cite{kenzelmann06,Cabrera} is shown in Fig.\ \ref{fig:cryst}(b). It should be noted that in these studies it was not possible to uniquely determine the phase relationship between the two sites. Recent polarised neutron diffraction measurements performed in an applied electric field $\mathbf{E}$ have demonstrated that the handedness of the cycloidal order can be switched by reversing $\mathbf{E}$ \cite{Cabrera}, although no information on the spatial distribution of  domains was obtained.

In this work we present the determination of the magnetic structure of Ni$_3$V$_2$O$_8$, refining the spin moments on the two nickel sites and their phase relationship, and a demonstration of the imaging of the magnetic domains and their electric field control.

\section{Experimental details}

\begin{figure}
\includegraphics[width=.4\textwidth]{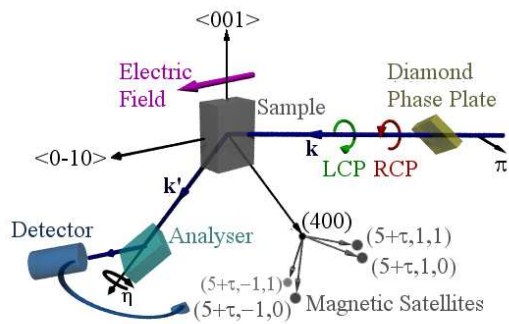}
\caption{Schematic of the experimental set up, where for a given magnetic satellite the Stokes parameters are determined by measuring the
scattered intensity as a function of the angle $\eta$.}\label{fig:expschem}
\end{figure}

NVO is orthorhombic with high temperature crystallographic structure $Cmce$ (number 64 in the International Crystallographic Tables) and lattice parameters $a=5.92$~\AA, $b=11.38$~\AA, and $c=8.24$~\AA.  The single crystal used for this experiment was grown from a BaO-V$_2$O$_5$ flux \cite{Lawes04}, and cut and mounted to give a specular $a$-face, with surface dimensions $600$~$\mu\mathrm{m}$$\times900$~$\mu\mathrm{m}$. Experiments were performed on the ID20 Magnetic Scattering beamline \cite{Paolasini07} at the European Synchrotron Radiation Facility in Grenoble, using a monochromatised X-ray beam at an energy of $7.45$~keV, considerably below the nickel $K$-edge to avoid interference with the resonant signal and to enable the separate determination of the spin and orbital components of the magnetic moments. The experimental setup is presented in Figure~\ref{fig:expschem}.

A polarisation analyser, assembled on the detector arm of a six circle diffractometer, was used to discriminate the scattered photon polarisation by using a high quality Au(222) analyzer with mosaic spread of $\sim0.22^\circ$. At this energy the cross-talk between the $\sigma^\prime$ (rotated, $\eta=0^\circ$) and $\pi^\prime$ (unrotated, $\eta=90^\circ$) channels was approximately $0.3\%$. This enables the extraction of the Stokes parameters by fitting the dependence of the scattered light $I = I_0 (1 + P_1\cos(2\eta) + P_2\sin(2\eta))$ on the angle $\eta$ of the analyser assembly about the scattering vector k \cite{deBergevin,Blume}.

An in-vacuum phase plate setup was used in quarter-wave mode to convert the incoming linearly polarised light ($\hat{\pi}$) into the circular left (LCP) and right (RCP) components, by using a high quality ($1\bar{1}0$) diamond single crystal plate ($1.2$~mm thick) in Laue geometry, and oriented at about $45^\circ$ with respect the incident polarisation $\hat{\pi}$, in order to have the (111) reflecting plane normal to the surface \cite{Scagnoli}.  The Stokes parameters of the incident beam were carefully checked before and after every set of scans by using the same polarisation analyser, in order to have a good control of the incident circular photon polarisation  (in this case $|P_1|\approx|P_2|\approx0.008(4)$, hence circular beam polarisation close to $99\%$, assuming no depolarisation of the beam such that $P_3=\sqrt(1-P_1^2-P_2^2)$), which is very sensitive to the beam position stability.

Measurements were performed by recording rocking curves of the analyser at a series of values of $\eta$. The intensity $I(\eta)$ was then obtained by numerically integrating the individual peak shapes for LCP and RCP incident. To correct our data for the diffuse Thomson scattering background, arising from the sigma component of the circular polarisation, measurements were also performed off the magnetic reflection ($\Delta\theta=0.2^\circ$) and then the background $I(\eta)$ was subtracted from the signal at the reflection.

An electric field of up to $2$~kV/mm was applied by means of two copper electrodes glued by high conductive silver paste on the sample $b$-surfaces. The sample stick was inserted into an orange cryostat allowing us to access a temperature range of $2-300$~K. NVO is an insulator, and in spite of the use of $^4$He exchange gas, given an incident photon flux of $\sim10^{13}$~ph/sec, it was necessary to attenuate the incident X-ray beam to reduce the transmission by a factor of five, in order to maintain the sample within the LTI phase.

\section{Experimental Results and Analysis}
\begin{figure}
\includegraphics[width=0.48\textwidth]{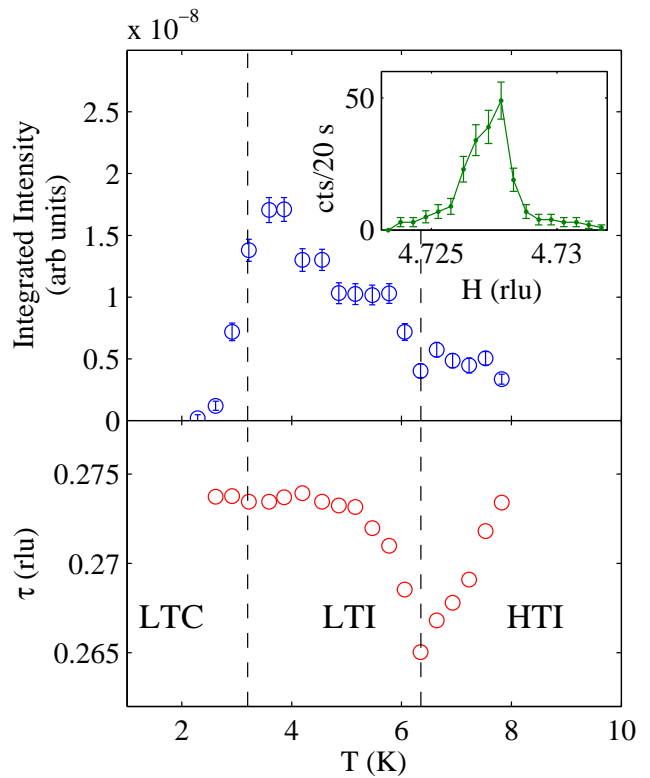}
\caption{Temperature dependence of the scattering intensity (upper) and wave-vector position (lower panel) of the magnetic satellite ($5-\tau$,1,0) measured in the $\pi-\sigma^\prime$ channel. The lines \cite{kenzelmann06} define the boundaries of the low temperature incommensurate phase of interest. The inset in the top panel shows a typical reciprocal space scan along the $h$-direction of the same satellite, measured at $T=4$~K.}\label{fig:tdep}
\end{figure}

Figure~\ref{fig:tdep} shows the temperature dependence of the magnetic reflection ($5-\tau$,1,0) obtained from scans measured in the rotated $\pi-\sigma$ channel along the reciprocal space $h$-direction. The sharp change seen in the intensity of the scattering at $T=3.2$~K, obtained by numerical integration, indicates the transition between the low temperature commensurate (LTC) and incommensurate (LTI) phases, whilst the discontinuity seen in the evolution of the propagation wave-vector at $T=6.3$~K occurs at the transition between the high (HTI) and low temperature incommensurate phases. Hence the LTI phase of interest was clearly defined, and our measurements were performed at an optimised temperature, taking great care to avoid thermal instabilities which could lead to crossing the phase boundaries.

\subsection{Magnetic Structure Refinement}
Our investigation of the magnetic structure was facilitated by the fact that an applied electric field can create a magnetic domain with a single handedness through the magneto-electric coupling \cite{Fabrizi}, and was accomplished by performing a full Stokes analysis of the scattered intensity, a technique which, as established for TbMnO$_3$ \cite{Fabrizi}, provides information on the magnetic structure that is not accessible to neutron diffraction experiments.  The field cooling procedure for performing the measurements was to apply $-1.4$~kV/mm across the sample as it was cooled from $20$~K, in the paramagnetic phase, to $4.2$~K, in the LTI phase. The electric field was then removed before illuminating the sample with X-rays.

\begin{figure}
\centering
\includegraphics[width=0.4\textwidth]{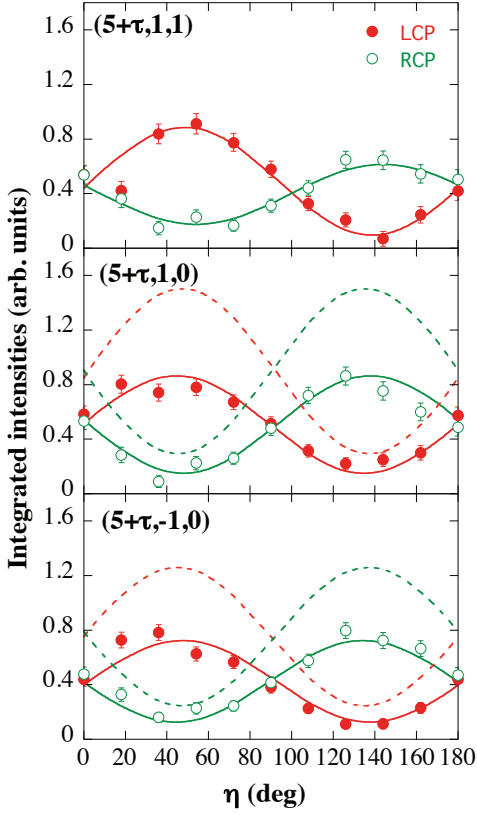}
\caption{The Stokes dependence of the scattered intensity from magnetic satellites on analyser rotation angle $\eta$ was determined by the data measured in the LTI phase with left (LCP) and right circularly polarised (RCP) X-rays after
field cooling with $E=1$~kV/mm along $-b$. The solid lines represent  calculations with spin moments of $m_a^{SP}$=1.6,  $m_b^{SP}$=1.3, $m_a^{CT}$=0, and $m_b^{CT}$=0, in units of $\mu_B$, and a domain population of
$87\%$. The dashed lines correspond to the neutron model \cite{kenzelmann06} with finite cross-tie moments (Fig.\ \ref{fig:cryst}(b)), with an assumed phase difference of $\pi/2$ between the spine and cross-tie moments,
and where the intensities are scaled to that of the upper panel. \label{fig:diffraction}}
\end{figure}

Figure~\ref{fig:diffraction} shows the Stokes dependence measured for three different reflections: ($5+\tau$, $1$, $1$), ($5+\tau$, $1$, $0$) and ($5+\tau$, $-1$, $0$) with both LCP and RCP X-rays incident, as red and green markers respectively. Attempts to observe a signal at a fourth reflection: ($4+\tau$, $1$, $0$) were unsuccessful. In all three panels one notices that the maxima and minima in the integrated intensities lie close to $45^\circ$ and $135^\circ$, where their positions are reversed for LCP vs RCP incident. In addition the large variation in the intensities as a function of $\eta$ indicates that the X-rays have been effectively converted on scattering from being circularly to linearly (in this case obliquely) polarised, i.e. $P_2>P_1>P_3$. This differs from the behaviour seen for TbMnO$_3$, in which the scattered X-rays retained a partially circular polarisation state \cite{Fabrizi}. For ($5+\tau$, $1$, $1$) in the first panel, the maximum intensity for LCP is greater than that for RCP, whilst in the lower two panels the maximum intensities are approximately equal for the two circular polarisations.

A particular feature of NVO is that by taking advantage of the structure factor one may selectively probe the spine and cross-tie moments independently or in combination. Magnetic reflections can be classified according to their indices ($hkl$): type 1 -- $h$ odd, $k$ odd, $l$ odd - cross-tie
moments cancel such that only the spine site moments contribute, type 2 -- $h$ even, $k$ even, $l$ odd - the reverse is true, such that the cross-tie moments are singled out, and type 3 -- $h$ odd, $k$ odd, $l$ even - both moment types contribute to the scattering. The three different peak type scattering amplitudes are:
\begin{eqnarray}
f_1&\propto&8\cos(2\pi k \Delta_B^{SP})(\mathbf{m}_a^{SP}-\mathrm{i}\gamma\alpha\mathbf{m}_b^{SP})\cdot\mathbf{B},\nonumber\\
f_2&\propto&4(\mathbf{m}_a^CT-\mathrm{i}\gamma\alpha\mathbf{m}_b^{CT})\cdot\mathbf{B},\nonumber\\
f_3&\propto&(8\beta\sin(2\pi k \Delta_B^{SP})(\mathbf{m}_a^{SP}-\mathrm{i}\gamma\alpha\mathbf{m}_b^{SP})\nonumber\\
&&+4\exp(\mathrm{i}\phi)(\mathbf{m}_a^{CT}-\mathrm{i}\gamma\alpha\mathbf{m}_b^{CT}))\cdot\mathbf{B},\label{eqn:fullSA}
\end{eqnarray}
where $\alpha=\pm1$ gives the sign of $\tau$ in the propagation wavevector, $\beta=(-1)^{n+p}$ for $h=2n+1$ and $l=2p$, $\gamma=\pm1$ for the two different domains, $\Delta_b^{SP}=0.13$ is the fractional coordinate along the $b$-axis of the Wyckoff position occupied by the spine ions, describing the buckling of the Kagom\'{e} planes, and the vector $\mathbf{B}$ is that of Blume and Gibbs\cite{Blume}, containing the polarisation state of the incident and scattered X-rays for the spin contribution. Thus the different magnetic reflections we investigated can be identified as : type 1 --($5+\tau$, $1$, $1$), type 2 -- ($4+\tau$, $1$, $0$) and type 3 -- ($5+\tau$, $+(-)1$, $0$), where the sign of $+(-)1$ enables us to extract information regarding the phase relationship $\phi$ between the two moment types.

Starting with the pure spine-type only reflection ($5+\tau$, $1$, $1$), we compare our data for both LCP and RCP with a calculation for the magnetic structure proposed in an earlier neutron diffraction study \cite{kenzelmann06}, with total magnetic moments:
\begin{eqnarray*}
m_a^{SP}&=&1.6\, \mu_B,\\
m_b^{SP}&=&1.3\, \mu_B.
\end{eqnarray*}
Our calculations use these total magnetic moments as spin moments only on the Ni ions, having assumed that the orbital moment is negligible\cite{comment}, and find an excellent agreement. Our fit, which is shown in the upper panel of Fig.~\ref{fig:diffraction} as a solid line, allowed us to extract the domain population which we found to be  $87(4)\%$ domain type 1, corresponding to the moments rotating clockwise in the $a-b$ plane as we move along the $a$ axis, when observed from $+c$.

The absence of an observed signal at the cross-tie site only reflection suggests that any moment on this site is very small. For the third type of reflection, exemplified by the twin reflections $(5+\tau, \pm 1, 0)$, to aid the understanding of the data and to investigate the cross-tie site moment and the phase relationship between the two moment sites, we can make use of a simplifying model.

We stress that the purpose of this simplified model is to allow us to understand the qualitative behaviour of the data shown in Fig.~\ref{fig:diffraction} only. The starting assumptions for the simplified model are: 1) a single domain, 2) an analyser Bragg angle of $\theta=45^\circ$, 3) the scattering plane is perfectly horizontal, and 4) within the scattering plane, the angle between the $b$-axis and the scattering vector is $45^\circ$, the scattering amplitudes in the $\hat{\sigma}^\prime$ and $\hat{\pi}^\prime$ channels can be written as:
\begin{eqnarray*}
f_{\hat{\sigma}^\prime}&\propto&[kS_a+C_b\sin\phi-C_a\cos\phi]\\
&&+\mathrm{i}[kS_b-C_b\cos\phi-C_a\sin\phi],\\
f_{\hat{\pi}^\prime}&\propto&[kS_b+C_b\cos\phi+C_a\sin\phi]\\
&&+\mathrm{i}\epsilon[kS_a-C_b\sin\phi-C_a\cos\phi],
\end{eqnarray*}
where
\begin{eqnarray*}
S_a&=&8\sin(2\pi\Delta_b^{SP})m_a^{SP},\\
S_b&=&8\sin(2\pi\Delta_b^{SP})m_b^{SP},\\
C_a&=&4m_a^{CT},\\
C_b&=&4m_b^{CT},
\end{eqnarray*}
$k=\pm1$ differentiates between the two reflections $(5+\tau,\,\pm1,\,0)$, and $\epsilon=\pm1$ refers to the handedness of the incident light. The average intensity for the two reflections can be expressed as:
\begin{eqnarray}
|f_{\hat{\sigma}^\prime}|^2+|f_{\hat{\pi}^\prime}|^2&=&|kS_a+C_b\sin\phi-C_a\cos\phi|^2\nonumber\\
&+&|kS_b-C_b\cos\phi-C_a\sin\phi|^2\nonumber\\
&+&|kS_b+C_b\cos\phi+C_a\sin\phi|^2\nonumber\\
&+&|kS_a-C_b\sin\phi-C_a\cos\phi|^2.\label{eqn:long}
\end{eqnarray}
It is noticeable that $\epsilon$, the sign of the incident polarisation, is now absent, consistent with the average intensity for LCP and RCP being the same. This arises from the spin moments lying in the scattering plane, in contrast to when measurements are made of the spine-type only reflection. If the observational constraint of equivalence between $\pm k$, i.e. between panels two and three of Fig.~\ref{fig:diffraction}, is applied to Eq.~\eqref{eqn:long} this requires that $C_a\cos\phi=0$ and $C_b\cos\phi=0$ and hence a phase difference of $\phi=\pi/2$. This results in further simplifying the scattering
amplitudes to:
\begin{eqnarray}
f_{\hat{\sigma}^\prime}&\propto&[kS_a+C_b]+\mathrm{i}[kS_b-C_a],\nonumber\\
f_{\hat{\pi}^\prime}&\propto&[kS_b+C_a]+\mathrm{i}\epsilon[kS_a-C_b].
\end{eqnarray}

If one then considers the observation that the scattering is predominantly described by $P_2$, with $P_1$ close to zero, where
\begin{eqnarray}
P_1&=&\frac{|f_{\hat{\sigma}^\prime}|^2-|f_{\hat{\pi}^\prime}|^2}{|f_{\hat{\sigma}^\prime}|^2+|f_{\hat{\pi}^\prime}|^2},\nonumber\\
P_2&=&\frac{|f_{\hat{\sigma}^\prime}+f_{\hat{\pi}^\prime}|^2-|f_{\hat{\sigma}^\prime}-f_{\hat{\pi}^\prime}|^2}{|f_{\hat{\sigma}^\prime}|^2+|f_{\hat{\pi}^\prime}|^2},
\end{eqnarray}
this imposes additional constraints. Firstly for $P_1\approx0$ this requires that $|f_{\hat{\sigma}^\prime}|\approx|f_{\hat{\pi}^\prime}|$, and hence that $S_a\approx S_b$ and $C_a\approx C_b$, which is broadly consistent with previously published values. Meanwhile $P_2$ is not equal to zero for
$|f_{\hat{\sigma}^\prime}+f_{\hat{\pi}^\prime}|\neq|f_{\hat{\sigma}^\prime}-f_{\hat{\pi}^\prime}|$,
expressions for which are:
\begin{widetext}
\begin{eqnarray}\label{eqn:P2}
\epsilon=+1:&|f_{\hat{\sigma}^\prime}+f_{\hat{\pi}^\prime}|^2&=|(S_a+S_b)+(C_a+C_b)|^2+|(S_a+S_b)-(C_a+C_b)|^2,\nonumber\\
&|f_{\hat{\sigma}^\prime}-f_{\hat{\pi}^\prime}|^2&=|(S_a-S_b)+(C_a-C_b)|^2+|(S_a-S_b)-(C_a-C_b)|^2,\nonumber\\
\epsilon=-1:&|f_{\hat{\sigma}^\prime}+f_{\hat{\pi}^\prime}|^2&=|(S_a-S_b)+(C_a-C_b)|^2+|(S_a-S_b)-(C_a-C_b)|^2,\nonumber\\
&|f_{\hat{\sigma}^\prime}-f_{\hat{\pi}^\prime}|^2&=|(S_a+S_b)+(C_a+C_b)|^2+|(S_a+S_b)-(C_a+C_b)|^2,
\end{eqnarray}
\end{widetext}
i.e. they are independent of the sign of $k$. Equation~\eqref{eqn:P2} clearly shows that reversing the handedness of the circular polarisation $\epsilon$ results in switching the sign of $P_2$. Since we have already established from $P_1\approx0$ that $S_a\approx S_b$ and $C_a\approx C_b$, then $(S_a-S_b)$ and $(C_a-C_b)$ are close to zero, whilst $|(S_a+S_b)+(C_a+C_b)|$ is strong, such that there is a high contrast between $|f_{\hat{\sigma}^\prime}+f_{\hat{\pi}^\prime}|$ and $|f_{\hat{\sigma}^\prime}-f_{\hat{\pi}^\prime}|$, resulting in large values for $P_2$, which is reversible depending on $\epsilon$, consistent with the key observations from the data shown in Figure~\ref{fig:diffraction}.

\begin{figure*}
\centering
\includegraphics[width=0.6\textwidth]{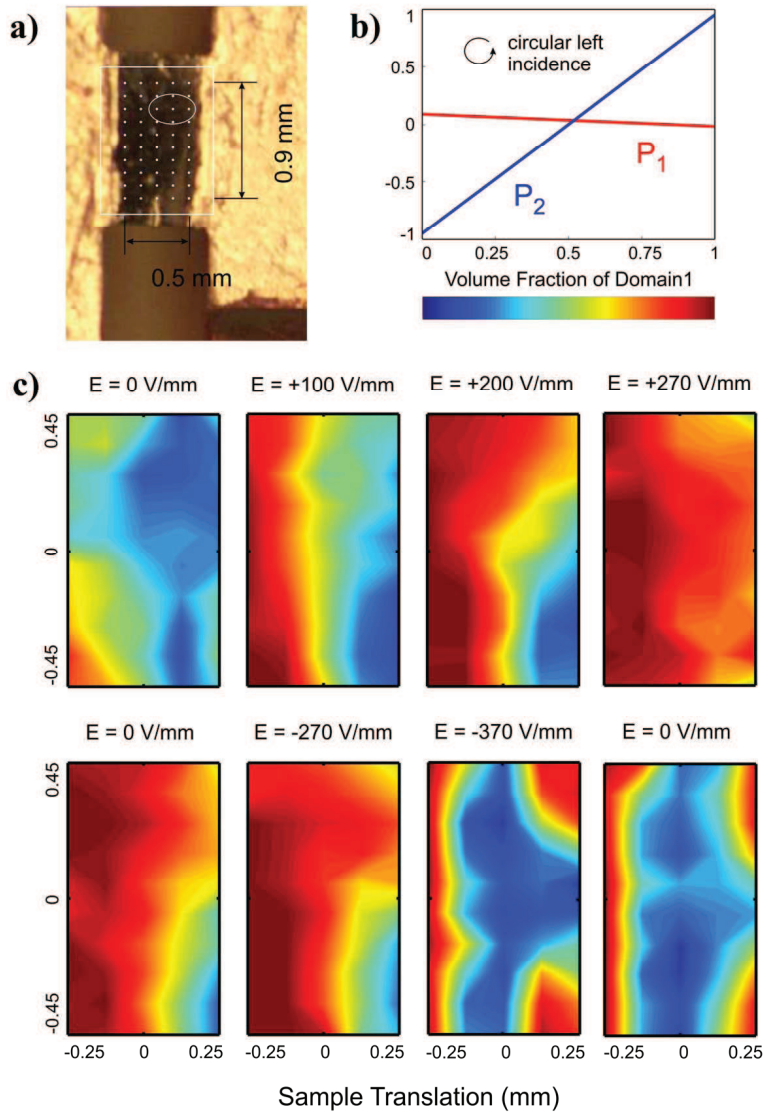}
\caption{(a) Image of the sample sandwiched between the Cu electrodes used to form a capacitor. The dotted mesh superimposed  on the image indicates points at which the domain population was determined.
The ellipse indicates the size of the X-ray beam.  (b) Variation of the Stokes parameters as a function of domain volume fraction, demonstrating the sensitivity to $P_2$ (blue line) and insensitivity to $P_1$ (red line), allowing a contrast to be measured between the two cycloidal domains. (c) Images of the domain populations as a function of applied electric field for the magnetic reflection ($5+\tau$, $1$, $0$), where the colour represents the percentage of the clockwise and anti-clockwise magnetic cycloidal domains.\label{fig:imaging}}
\end{figure*}

\begin{figure}
\centering
\includegraphics[width=0.45\textwidth]{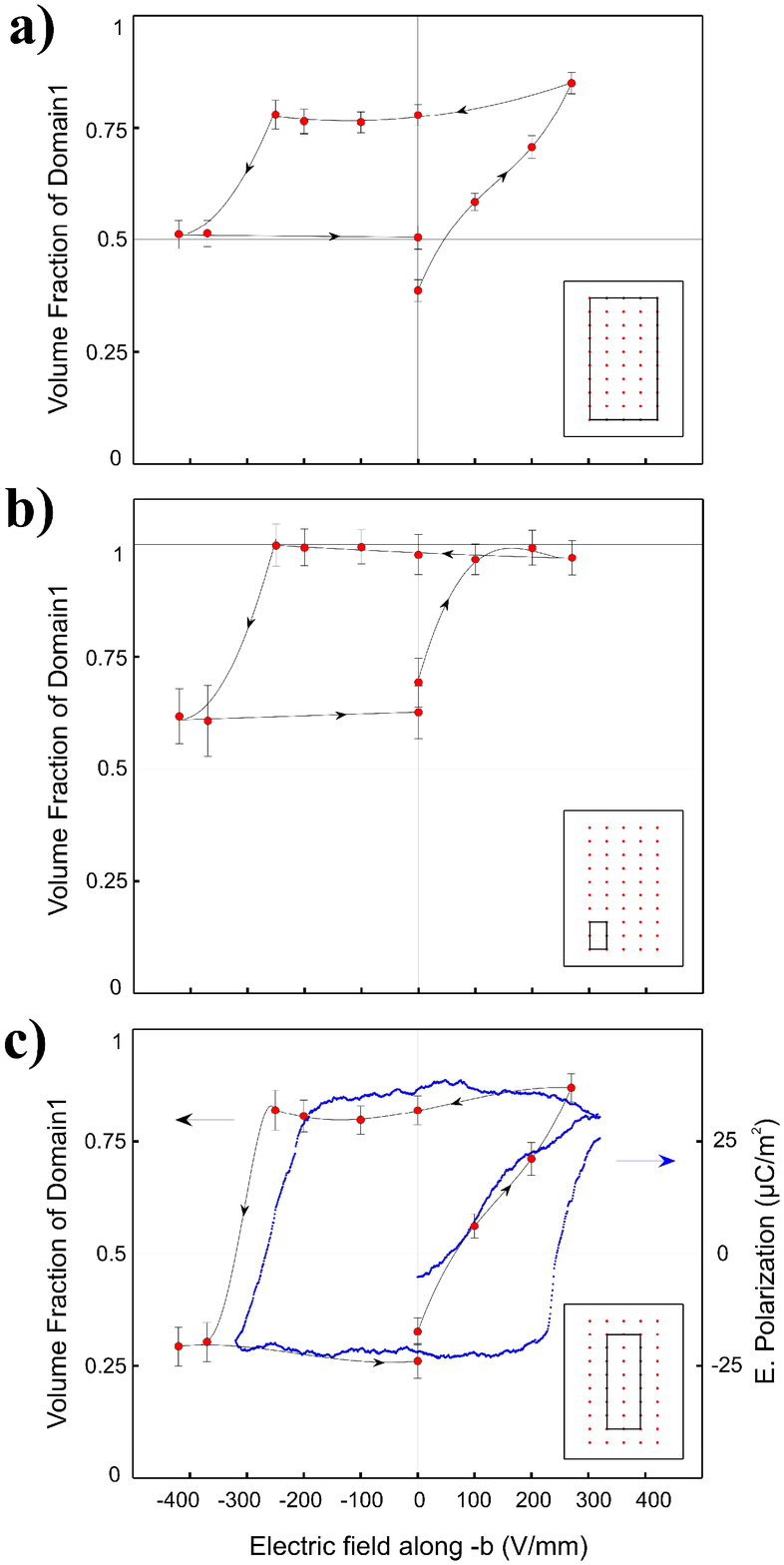}
\caption{Position dependent magnetic domain population hysteresis loops obtained from the domain images shown in Fig.\ \ref{fig:imaging} averaged over different areas of interest: (a) entire sample area; (b) edge; (c) central region.  In (c)
the data are compared with bulk measurements of the electric polarisation \cite{Cabrera} where an appropriate scaling has been applied. \label{fig:hystloop}}
\end{figure}

Using the moment magnitudes and intensity scaling factor extracted for the spine-type sites given above, a fit was made simultaneously to the different data for the twin reflections using the full model given in Eq.~\eqref{eqn:fullSA}, including the full six-circle geometry, and this is shown by the solid lines in the lower two panels of Figure~\ref{fig:diffraction}. The scattered signal is found to be very sensitive to the value of the phase difference between the moments on the spine and cross-tie sites, with the similarity in the scattered intensities for the two reflections implying a phase difference of 0.50(2)$\pi$. Calculations for different magnetic structure models allow us to provide estimates of $m^{CT}_a$ = $0.55(50)$ and $m^{CT}_b$= $0.35(35)$ (units of $\mu_B$), with a 87(4) \% domain type 1 population, where the large error bars reflect the relative insensitivity to the cross-tie moments. Nonetheless, the data and model calculations displayed in Fig.~\ref{fig:diffraction} underline the remarkable sensitivity of the technique to the magnetic amplitudes and phase relationships of  complex magnetic structures. It is noteworthy that the observed cross-tie moments are considerably smaller than the published values obtained from neutron diffraction \cite{Cabrera} (calculations for which are shown in Fig.~\ref{fig:diffraction} as dashed lines), although the analysis of more recent neutron diffraction measurements seems consistent with the possibility of there being no moment on the cross-tie nickel ions [I. Cabrera et al, unpublished]. Thus we may conclude that the cross-tie moments are most likely negligible, from which it follows that spatial inversion symmetry is broken by the non-collinear magnetic structure of the spine-sites alone.

\subsection{Domain Imaging}
The data in Fig.\ \ref{fig:diffraction} also suggest a novel route to imaging multiferroic domains. The reversal evident in the sign of the Stokes parameter $P_2$ when switching the incident X-ray polarization, would also occur for fixed incident polarization if the cycloidal magnetic domain
switched its handedness \cite{Fabrizi}. It follows that the domain population can therefore  be determined at different points in the sample
by determining the value of $P_2$ (see Fig.\ \ref{fig:imaging}(b)). ($P_1$ and $P_3$ are relatively insensitive to domain volume fraction.) For the imaging the incident beam spot was reduced to $0.25\times0.38$~mm$^2$ to maximise the resolution whilst still giving a reasonable count rate in the detector (Fig.\ \ref{fig:imaging}(a)). The sample was then rastered through the beam and at each point, the scattered intensity was recorded for the analyser setting angles of $\eta=45^\circ$ and $135^\circ$. The Stokes parameter $P_2$ was evaluated from these measurements, hence providing the percentage of  left or right handed magnetic cycloidal domains. To establish the evolution of the domains as a function of electric field a different field application protocol was employed. Initially the sample was cooled to $5$~K with no voltage across the sample. Then between each set of measurements, whilst remaining at $5$~K, the voltage across the sample was ramped up to the required level. Images of the domains are shown in Fig.\ \ref{fig:imaging}(c) as a function of electric field applied, demonstrating clearly that we were able to resolve inhomogeneities in the domain populations with good resolution.

Inspection of the individual images reveals that for zero applied electric field the distribution of the two domains is not equal, but instead shows a preference towards domain 2, i.e. moments rotating anti-clockwise. This may either be a surface strain effect or some memory effect of an earlier domain state induced by field cooling in previous experiments, as has been observed in MnWO$_4$ \cite{Finger}. Applying an increasing electric field along $-b$ leads to a reversal of the domain populations, with the production of almost a mono-domain 1 state for $E=+270$~V/mm. The evolution of the domain populations is gradual, with the boundary between the two shifting as a function of applied electric field, as opposed to the nucleation of domain 1 within domain 2 leading to a more randomised distribution. When the electric field direction is then reversed the domain distributions are little changed  until between $-250$~V/mm and $-370$~V/mm the distribution is again reversed with a preference towards domain 2. Whilst the domain populations are preserved on returning to $E=0$~V/mm, due to time constraints no additional data were collected for applied positive electric fields.

One way to quantitatively render the domain homogeneities is to average the images over different areas of interest to produce position dependent magnetic domain population hysteresis loops. Figure \ref{fig:hystloop}(a) shows the domain population averaged over the entire sample. On average the excess population of domain 2 in zero field is quickly reversed in a positive applied field, with domain 1 prevailing such that even large negative fields are incapable of restoring an excess of domain 2. In Fig.\  \ref{fig:hystloop}(b), the response from the bottom left-hand corner is plotted, where it is clear that close to the edge of the sample there is strong pinning of domain 1 with it dominating at all fields. Imaging allows these edge effects to be excluded from the data, (Fig.\ \ref{fig:imaging}(c)), and restores more of a symmetrical character to the electric field dependence of the magnetic domain population. This allows us to compare our data  with the electric polarisation measurements of Cabrera et al \cite{Cabrera},  as shown in Fig.\ \ref{fig:hystloop}(c), where we find excellent agreement. This establishes the link between the magnetic domains and the electric polarisation in multiferroic NVO, proving that by imaging the magnetic domains we are in effect also imaging the ferroelectric domains.

\section{Conclusion}
In conclusion, we have refined the magnetic structure of Ni$_3$V$_2$O$_8$, confirming the cycloidal ordering of the Ni spins on the spine sites whilst finding that there is no ordered moment on the cross-tie sites, indicating that it is the magnetic order on the spine sites alone that breaks inversion symmetry. Further, we have successfully demonstrated how polarisation enhanced X-ray imaging (PEXI), through the contrast provided by the magnetic structure, enables the imaging of the magnetic domains in Ni$_3$V$_2$O$_8$ which endow this material with its multiferroic properties.

Comparison of the magnetic domain population hysteresis loops with bulk electric polarisation measurements reveals the coupling between the magnetic and ferroelectric domains. This opens the prospect of using this technique to image multiferroic domains in related materials, and even in operational devices. With the improved focussing offered by beamlines currently under development the spatial resolution for this technique will be reduced to 100 nm and below, at which point it will be possible not only to image the domains themselves, but also the structure of the domain walls, the engineering of which ultimately determines the usefulness of a given material.

Many thanks to everyone who has helped us with the experiments,
especially A. Fondacaro, J. Herrero-Martin, C. Mazzoli, V. Scagnoli and A. C. Walters.

\bibliography{nivo}

\begin{thebibliography}{36}
\expandafter\ifx\csname natexlab\endcsname\relax\def\natexlab#1{#1}\fi
\expandafter\ifx\csname bibnamefont\endcsname\relax
  \def\bibnamefont#1{#1}\fi
\expandafter\ifx\csname bibfnamefont\endcsname\relax
  \def\bibfnamefont#1{#1}\fi
\expandafter\ifx\csname citenamefont\endcsname\relax
  \def\citenamefont#1{#1}\fi
\expandafter\ifx\csname url\endcsname\relax
  \def\url#1{\texttt{#1}}\fi
\expandafter\ifx\csname urlprefix\endcsname\relax\def\urlprefix{URL }\fi
\providecommand{\bibinfo}[2]{#2}
\providecommand{\eprint}[2][]{\url{#2}}

\bibitem[{\citenamefont{David et~al.}(2002)\citenamefont{David, N\"ohammer, and
  Solak}}]{David02}
\bibinfo{author}{\bibfnamefont{C.}~\bibnamefont{David}},
  \bibinfo{author}{\bibfnamefont{B.}~\bibnamefont{N\"ohammer}},
  \bibnamefont{and} \bibinfo{author}{\bibfnamefont{H.~H.} \bibnamefont{Solak}},
  \bibinfo{journal}{Appl. Phys. Lett.} \textbf{\bibinfo{volume}{81}},
  \bibinfo{pages}{3287} (\bibinfo{year}{2002}).

\bibitem[{\citenamefont{Weitkamp et~al.}(2005)\citenamefont{Weitkamp,
  N\"ohammer, Diaz, David, and Ziegler}}]{Weitkamp05}
\bibinfo{author}{\bibfnamefont{T.}~\bibnamefont{Weitkamp}},
  \bibinfo{author}{\bibfnamefont{B.}~\bibnamefont{N\"ohammer}},
  \bibinfo{author}{\bibfnamefont{A.}~\bibnamefont{Diaz}},
  \bibinfo{author}{\bibfnamefont{C.}~\bibnamefont{David}}, \bibnamefont{and}
  \bibinfo{author}{\bibfnamefont{E.}~\bibnamefont{Ziegler}},
  \bibinfo{journal}{Appl. Phys. Lett.} \textbf{\bibinfo{volume}{86}},
  \bibinfo{pages}{054101} (\bibinfo{year}{2005}).

\bibitem[{\citenamefont{Pfeiffer et~al.}(2006)\citenamefont{Pfeiffer, Weitkamp,
  Bunk, and David}}]{Pfeiffer08}
\bibinfo{author}{\bibfnamefont{F.}~\bibnamefont{Pfeiffer}},
  \bibinfo{author}{\bibfnamefont{T.}~\bibnamefont{Weitkamp}},
  \bibinfo{author}{\bibfnamefont{O.}~\bibnamefont{Bunk}}, \bibnamefont{and}
  \bibinfo{author}{\bibfnamefont{C.}~\bibnamefont{David}},
  \bibinfo{journal}{Nature Physics} \textbf{\bibinfo{volume}{2}},
  \bibinfo{pages}{268} (\bibinfo{year}{2006}).

\bibitem[{\citenamefont{Miao et~al.}(1999)\citenamefont{Miao, Charalambous,
  Kirz, and Sayre}}]{Miao99}
\bibinfo{author}{\bibfnamefont{J.}~\bibnamefont{Miao}},
  \bibinfo{author}{\bibfnamefont{P.}~\bibnamefont{Charalambous}},
  \bibinfo{author}{\bibfnamefont{J.}~\bibnamefont{Kirz}}, \bibnamefont{and}
  \bibinfo{author}{\bibfnamefont{D.}~\bibnamefont{Sayre}},
  \bibinfo{journal}{Nature} \textbf{\bibinfo{volume}{400}},
  \bibinfo{pages}{342} (\bibinfo{year}{1999}).

\bibitem[{\citenamefont{Robinson et~al.}(2001)\citenamefont{Robinson,
  Vartanyants, Williams, Pfeifer, and Pitney}}]{Robinson01}
\bibinfo{author}{\bibfnamefont{I.~K.} \bibnamefont{Robinson}},
  \bibinfo{author}{\bibfnamefont{I.~A.} \bibnamefont{Vartanyants}},
  \bibinfo{author}{\bibfnamefont{G.~J.} \bibnamefont{Williams}},
  \bibinfo{author}{\bibfnamefont{M.~A.} \bibnamefont{Pfeifer}},
  \bibnamefont{and} \bibinfo{author}{\bibfnamefont{J.~A.}
  \bibnamefont{Pitney}}, \bibinfo{journal}{Phys. Rev. Lett.}
  \textbf{\bibinfo{volume}{87}}, \bibinfo{pages}{195505}
  (\bibinfo{year}{2001}).

\bibitem[{\citenamefont{McNulty et~al.}(1992)\citenamefont{McNulty, Kirz,
  Jacobsen, Anderson, Howells, and Kern}}]{McNulty92}
\bibinfo{author}{\bibfnamefont{I.}~\bibnamefont{McNulty}},
  \bibinfo{author}{\bibfnamefont{J.}~\bibnamefont{Kirz}},
  \bibinfo{author}{\bibfnamefont{C.}~\bibnamefont{Jacobsen}},
  \bibinfo{author}{\bibfnamefont{E.}~\bibnamefont{Anderson}},
  \bibinfo{author}{\bibfnamefont{M.}~\bibnamefont{Howells}}, \bibnamefont{and}
  \bibinfo{author}{\bibfnamefont{D.}~\bibnamefont{Kern}},
  \bibinfo{journal}{Science} \textbf{\bibinfo{volume}{256}},
  \bibinfo{pages}{1009} (\bibinfo{year}{1992}).

\bibitem[{\citenamefont{Eisebitt et~al.}(2004)\citenamefont{Eisebitt, L\"uning,
  Schlotter, L\"orgen, Hellwig, Eberhardt, and St\"ohr}}]{Eisebitt04}
\bibinfo{author}{\bibfnamefont{S.}~\bibnamefont{Eisebitt}},
  \bibinfo{author}{\bibfnamefont{J.}~\bibnamefont{L\"uning}},
  \bibinfo{author}{\bibfnamefont{W.}~\bibnamefont{Schlotter}},
  \bibinfo{author}{\bibfnamefont{M.}~\bibnamefont{L\"orgen}},
  \bibinfo{author}{\bibfnamefont{O.}~\bibnamefont{Hellwig}},
  \bibinfo{author}{\bibfnamefont{W.}~\bibnamefont{Eberhardt}},
  \bibnamefont{and} \bibinfo{author}{\bibfnamefont{J.}~\bibnamefont{St\"ohr}},
  \bibinfo{journal}{Nature} \textbf{\bibinfo{volume}{432}},
  \bibinfo{pages}{885} (\bibinfo{year}{2004}).

\bibitem[{\citenamefont{Kimura et~al.}(2003)\citenamefont{Kimura, Goto,
  Shintani, Ishizaka, Arima, and Tokura}}]{KimuraTMO}
\bibinfo{author}{\bibfnamefont{T.}~\bibnamefont{Kimura}},
  \bibinfo{author}{\bibfnamefont{T.}~\bibnamefont{Goto}},
  \bibinfo{author}{\bibfnamefont{H.}~\bibnamefont{Shintani}},
  \bibinfo{author}{\bibfnamefont{K.}~\bibnamefont{Ishizaka}},
  \bibinfo{author}{\bibfnamefont{T.}~\bibnamefont{Arima}}, \bibnamefont{and}
  \bibinfo{author}{\bibfnamefont{Y.}~\bibnamefont{Tokura}},
  \bibinfo{journal}{Nature} \textbf{\bibinfo{volume}{426}}, \bibinfo{pages}{55}
  (\bibinfo{year}{2003}).

\bibitem[{\citenamefont{Hur et~al.}(2004)\citenamefont{Hur, Park, Sharma, Ahn,
  Guha, and Cheong}}]{Hur}
\bibinfo{author}{\bibfnamefont{N.}~\bibnamefont{Hur}},
  \bibinfo{author}{\bibfnamefont{S.}~\bibnamefont{Park}},
  \bibinfo{author}{\bibfnamefont{P.~A.} \bibnamefont{Sharma}},
  \bibinfo{author}{\bibfnamefont{J.~S.} \bibnamefont{Ahn}},
  \bibinfo{author}{\bibfnamefont{S.}~\bibnamefont{Guha}}, \bibnamefont{and}
  \bibinfo{author}{\bibfnamefont{S.-W.} \bibnamefont{Cheong}},
  \bibinfo{journal}{Nature} \textbf{\bibinfo{volume}{429}},
  \bibinfo{pages}{392} (\bibinfo{year}{2004}).

\bibitem[{\citenamefont{Heyer et~al.}(2006)\citenamefont{Heyer, Hollmann,
  Klassen, Jodlauk, Bohat\'{y}, Becker, Mydosh, Lorenz, and Khomskii}}]{heyer}
\bibinfo{author}{\bibfnamefont{O.}~\bibnamefont{Heyer}},
  \bibinfo{author}{\bibfnamefont{N.}~\bibnamefont{Hollmann}},
  \bibinfo{author}{\bibfnamefont{I.}~\bibnamefont{Klassen}},
  \bibinfo{author}{\bibfnamefont{S.}~\bibnamefont{Jodlauk}},
  \bibinfo{author}{\bibfnamefont{L.}~\bibnamefont{Bohat\'{y}}},
  \bibinfo{author}{\bibfnamefont{P.}~\bibnamefont{Becker}},
  \bibinfo{author}{\bibfnamefont{J.~A.} \bibnamefont{Mydosh}},
  \bibinfo{author}{\bibfnamefont{T.}~\bibnamefont{Lorenz}}, \bibnamefont{and}
  \bibinfo{author}{\bibfnamefont{D.}~\bibnamefont{Khomskii}},
  \bibinfo{journal}{J. Phys. Condens. Matter} \textbf{\bibinfo{volume}{18}},
  \bibinfo{pages}{L471} (\bibinfo{year}{2006}).

\bibitem[{\citenamefont{Fiebig et~al.}(2002)\citenamefont{Fiebig, Lottermoser,
  Fr\"{o}hlich, Goltsev, and Pisarev}}]{Fiebignat}
\bibinfo{author}{\bibfnamefont{M.}~\bibnamefont{Fiebig}},
  \bibinfo{author}{\bibfnamefont{T.}~\bibnamefont{Lottermoser}},
  \bibinfo{author}{\bibfnamefont{D.}~\bibnamefont{Fr\"{o}hlich}},
  \bibinfo{author}{\bibfnamefont{A.~V.} \bibnamefont{Goltsev}},
  \bibnamefont{and} \bibinfo{author}{\bibfnamefont{R.~V.}
  \bibnamefont{Pisarev}}, \bibinfo{journal}{Nature}
  \textbf{\bibinfo{volume}{419}}, \bibinfo{pages}{818} (\bibinfo{year}{2002}).

\bibitem[{\citenamefont{Schierle et~al.}()\citenamefont{Schierle, Soltwisch,
  Schmitz, Feyerherm, Maljuk, Yokaichiya, Argyriou, and Weschke}}]{Schierle}
\bibinfo{author}{\bibfnamefont{E.}~\bibnamefont{Schierle}},
  \bibinfo{author}{\bibfnamefont{V.}~\bibnamefont{Soltwisch}},
  \bibinfo{author}{\bibfnamefont{D.}~\bibnamefont{Schmitz}},
  \bibinfo{author}{\bibfnamefont{R.}~\bibnamefont{Feyerherm}},
  \bibinfo{author}{\bibfnamefont{A.}~\bibnamefont{Maljuk}},
  \bibinfo{author}{\bibfnamefont{F.}~\bibnamefont{Yokaichiya}},
  \bibinfo{author}{\bibfnamefont{D.~N.} \bibnamefont{Argyriou}},
  \bibnamefont{and} \bibinfo{author}{\bibfnamefont{E.}~\bibnamefont{Weschke}},
  \bibinfo{note}{cond-mat 0910.5663}.

\bibitem[{\citenamefont{Rejm\'{a}nkov\'{a}
  et~al.}(1998)\citenamefont{Rejm\'{a}nkov\'{a}, Baruchel, Moretti, Arbore,
  Fejer, and Foulon}}]{Rejmankova}
\bibinfo{author}{\bibfnamefont{P.}~\bibnamefont{Rejm\'{a}nkov\'{a}}},
  \bibinfo{author}{\bibfnamefont{J.}~\bibnamefont{Baruchel}},
  \bibinfo{author}{\bibfnamefont{P.}~\bibnamefont{Moretti}},
  \bibinfo{author}{\bibfnamefont{M.}~\bibnamefont{Arbore}},
  \bibinfo{author}{\bibfnamefont{M.}~\bibnamefont{Fejer}}, \bibnamefont{and}
  \bibinfo{author}{\bibfnamefont{G.}~\bibnamefont{Foulon}},
  \bibinfo{journal}{J. Appl. Cryst.} \textbf{\bibinfo{volume}{31}},
  \bibinfo{pages}{106} (\bibinfo{year}{1998}).

\bibitem[{\citenamefont{Gonz\'{a}lez-Manas
  et~al.}(2005)\citenamefont{Gonz\'{a}lez-Manas, Vallejo, and
  Caballero}}]{Gonzalez}
\bibinfo{author}{\bibfnamefont{M.}~\bibnamefont{Gonz\'{a}lez-Manas}},
  \bibinfo{author}{\bibfnamefont{B.}~\bibnamefont{Vallejo}}, \bibnamefont{and}
  \bibinfo{author}{\bibfnamefont{M.~A.} \bibnamefont{Caballero}},
  \bibinfo{journal}{J. Appl. Cryst.} \textbf{\bibinfo{volume}{38}},
  \bibinfo{pages}{1012} (\bibinfo{year}{2005}).

\bibitem[{\citenamefont{Hong et~al.}(2001)\citenamefont{Hong, Woo, Shin, Jeon,
  pak, Colla, Setter, Kim, and No}}]{Hong}
\bibinfo{author}{\bibfnamefont{S.}~\bibnamefont{Hong}},
  \bibinfo{author}{\bibfnamefont{J.}~\bibnamefont{Woo}},
  \bibinfo{author}{\bibfnamefont{H.}~\bibnamefont{Shin}},
  \bibinfo{author}{\bibfnamefont{J.~U.} \bibnamefont{Jeon}},
  \bibinfo{author}{\bibfnamefont{Y.~E.} \bibnamefont{pak}},
  \bibinfo{author}{\bibfnamefont{E.~L.} \bibnamefont{Colla}},
  \bibinfo{author}{\bibfnamefont{N.}~\bibnamefont{Setter}},
  \bibinfo{author}{\bibfnamefont{E.}~\bibnamefont{Kim}}, \bibnamefont{and}
  \bibinfo{author}{\bibfnamefont{K.}~\bibnamefont{No}}, \bibinfo{journal}{J.
  Appl. Phys.} \textbf{\bibinfo{volume}{89}}, \bibinfo{pages}{1377}
  (\bibinfo{year}{2001}).

\bibitem[{\citenamefont{Nolting et~al.}(2000)\citenamefont{Nolting, Scholl,
  St\"ohr, Seo, Fompeyrine, Siegwart, Locquet, Anders, ning, Fullerton
  et~al.}}]{Nolting}
\bibinfo{author}{\bibfnamefont{F.}~\bibnamefont{Nolting}},
  \bibinfo{author}{\bibfnamefont{A.}~\bibnamefont{Scholl}},
  \bibinfo{author}{\bibfnamefont{J.}~\bibnamefont{St\"ohr}},
  \bibinfo{author}{\bibfnamefont{J.~W.} \bibnamefont{Seo}},
  \bibinfo{author}{\bibfnamefont{J.}~\bibnamefont{Fompeyrine}},
  \bibinfo{author}{\bibfnamefont{H.}~\bibnamefont{Siegwart}},
  \bibinfo{author}{\bibfnamefont{J.-P.} \bibnamefont{Locquet}},
  \bibinfo{author}{\bibfnamefont{S.}~\bibnamefont{Anders}},
  \bibinfo{author}{\bibfnamefont{J.~L.} \bibnamefont{ning}},
  \bibinfo{author}{\bibfnamefont{E.~E.} \bibnamefont{Fullerton}},
  \bibnamefont{et~al.}, \bibinfo{journal}{Nature}
  \textbf{\bibinfo{volume}{405}}, \bibinfo{pages}{767} (\bibinfo{year}{2000}).

\bibitem[{\citenamefont{Evans et~al.}(2002)\citenamefont{Evans, Isaacs, Aeppli,
  Cai, and Lai}}]{Evans}
\bibinfo{author}{\bibfnamefont{P.}~\bibnamefont{Evans}},
  \bibinfo{author}{\bibfnamefont{E.~D.} \bibnamefont{Isaacs}},
  \bibinfo{author}{\bibfnamefont{G.}~\bibnamefont{Aeppli}},
  \bibinfo{author}{\bibfnamefont{Z.}~\bibnamefont{Cai}}, \bibnamefont{and}
  \bibinfo{author}{\bibfnamefont{B.}~\bibnamefont{Lai}},
  \bibinfo{journal}{Science} \textbf{\bibinfo{volume}{295}},
  \bibinfo{pages}{1042} (\bibinfo{year}{2002}).

\bibitem[{\citenamefont{Shpyrko et~al.}(2007)\citenamefont{Shpyrko, Isaacs,
  Logan, Feng, Aeppli, Jaramillo, Kim, Rosenbaum, Zschack, Sprung
  et~al.}}]{Shpyrko}
\bibinfo{author}{\bibfnamefont{O.~G.} \bibnamefont{Shpyrko}},
  \bibinfo{author}{\bibfnamefont{E.~D.} \bibnamefont{Isaacs}},
  \bibinfo{author}{\bibfnamefont{J.~M.} \bibnamefont{Logan}},
  \bibinfo{author}{\bibfnamefont{Y.}~\bibnamefont{Feng}},
  \bibinfo{author}{\bibfnamefont{G.}~\bibnamefont{Aeppli}},
  \bibinfo{author}{\bibfnamefont{R.}~\bibnamefont{Jaramillo}},
  \bibinfo{author}{\bibfnamefont{H.~C.} \bibnamefont{Kim}},
  \bibinfo{author}{\bibfnamefont{T.~F.} \bibnamefont{Rosenbaum}},
  \bibinfo{author}{\bibfnamefont{P.}~\bibnamefont{Zschack}},
  \bibinfo{author}{\bibfnamefont{M.}~\bibnamefont{Sprung}},
  \bibnamefont{et~al.}, \bibinfo{journal}{Nature}
  \textbf{\bibinfo{volume}{68}}, \bibinfo{pages}{447} (\bibinfo{year}{2007}).

\bibitem[{\citenamefont{Kenzelmann et~al.}(2006)\citenamefont{Kenzelmann,
  Harris, Aharony, Entin-Wohlman, Yildirim, Huang, Park, Lawes, Broholm, Rogado
  et~al.}}]{kenzelmann06}
\bibinfo{author}{\bibfnamefont{M.}~\bibnamefont{Kenzelmann}},
  \bibinfo{author}{\bibfnamefont{A.~B.} \bibnamefont{Harris}},
  \bibinfo{author}{\bibfnamefont{A.}~\bibnamefont{Aharony}},
  \bibinfo{author}{\bibfnamefont{O.}~\bibnamefont{Entin-Wohlman}},
  \bibinfo{author}{\bibfnamefont{T.}~\bibnamefont{Yildirim}},
  \bibinfo{author}{\bibfnamefont{Q.}~\bibnamefont{Huang}},
  \bibinfo{author}{\bibfnamefont{S.}~\bibnamefont{Park}},
  \bibinfo{author}{\bibfnamefont{G.}~\bibnamefont{Lawes}},
  \bibinfo{author}{\bibfnamefont{C.}~\bibnamefont{Broholm}},
  \bibinfo{author}{\bibfnamefont{N.}~\bibnamefont{Rogado}},
  \bibnamefont{et~al.}, \bibinfo{journal}{Phys. Rev. B}
  \textbf{\bibinfo{volume}{74}}, \bibinfo{pages}{014429}
  (\bibinfo{year}{2006}).

\bibitem[{\citenamefont{Beutier et~al.}(2008)\citenamefont{Beutier, Bombardi,
  Vecchini, Radaelli, Park, Cheong, and Chapon}}]{Beutier}
\bibinfo{author}{\bibfnamefont{G.}~\bibnamefont{Beutier}},
  \bibinfo{author}{\bibfnamefont{A.}~\bibnamefont{Bombardi}},
  \bibinfo{author}{\bibfnamefont{C.}~\bibnamefont{Vecchini}},
  \bibinfo{author}{\bibfnamefont{P.~G.} \bibnamefont{Radaelli}},
  \bibinfo{author}{\bibfnamefont{S.}~\bibnamefont{Park}},
  \bibinfo{author}{\bibfnamefont{S.-W.} \bibnamefont{Cheong}},
  \bibnamefont{and} \bibinfo{author}{\bibfnamefont{L.~C.}
  \bibnamefont{Chapon}}, \bibinfo{journal}{Phys. Rev. B}
  \textbf{\bibinfo{volume}{77}}, \bibinfo{pages}{172408}
  (\bibinfo{year}{2008}).

\bibitem[{\citenamefont{Ewings et~al.}(2008)\citenamefont{Ewings, Boothroyd,
  McMorrow, Mannix, Walker, and Wanklyn}}]{Ewings}
\bibinfo{author}{\bibfnamefont{R.~A.} \bibnamefont{Ewings}},
  \bibinfo{author}{\bibfnamefont{A.~T.} \bibnamefont{Boothroyd}},
  \bibinfo{author}{\bibfnamefont{D.~F.} \bibnamefont{McMorrow}},
  \bibinfo{author}{\bibfnamefont{D.}~\bibnamefont{Mannix}},
  \bibinfo{author}{\bibfnamefont{H.~C.} \bibnamefont{Walker}},
  \bibnamefont{and} \bibinfo{author}{\bibfnamefont{B.~M.~R.}
  \bibnamefont{Wanklyn}}, \bibinfo{journal}{Phys. Rev. B}
  \textbf{\bibinfo{volume}{77}}, \bibinfo{pages}{104415}
  (\bibinfo{year}{2008}).

\bibitem[{\citenamefont{Feyerherm et~al.}(2009)\citenamefont{Feyerherm, Dudzik,
  Wolter, Valencia, Prokhnenko, Maljuk, Landsgesell, Aliouane, Bouchenoire,
  Brown et~al.}}]{Feyerherm}
\bibinfo{author}{\bibfnamefont{R.}~\bibnamefont{Feyerherm}},
  \bibinfo{author}{\bibfnamefont{E.}~\bibnamefont{Dudzik}},
  \bibinfo{author}{\bibfnamefont{A.~U.~B.} \bibnamefont{Wolter}},
  \bibinfo{author}{\bibfnamefont{S.}~\bibnamefont{Valencia}},
  \bibinfo{author}{\bibfnamefont{O.}~\bibnamefont{Prokhnenko}},
  \bibinfo{author}{\bibfnamefont{A.}~\bibnamefont{Maljuk}},
  \bibinfo{author}{\bibfnamefont{S.}~\bibnamefont{Landsgesell}},
  \bibinfo{author}{\bibfnamefont{N.}~\bibnamefont{Aliouane}},
  \bibinfo{author}{\bibfnamefont{L.}~\bibnamefont{Bouchenoire}},
  \bibinfo{author}{\bibfnamefont{S.}~\bibnamefont{Brown}},
  \bibnamefont{et~al.}, \bibinfo{journal}{Phys. Rev. B}
  \textbf{\bibinfo{volume}{79}}, \bibinfo{pages}{134426}
  (\bibinfo{year}{2009}).

\bibitem[{\citenamefont{Mannix et~al.}(2007)\citenamefont{Mannix, McMorrow,
  Ewings, Boothroyd, Prabhakaran, Joly, Janousova, Mazzoli, Paolasini, and
  Wilkins}}]{mannix}
\bibinfo{author}{\bibfnamefont{D.}~\bibnamefont{Mannix}},
  \bibinfo{author}{\bibfnamefont{D.~F.} \bibnamefont{McMorrow}},
  \bibinfo{author}{\bibfnamefont{R.~A.} \bibnamefont{Ewings}},
  \bibinfo{author}{\bibfnamefont{A.~T.} \bibnamefont{Boothroyd}},
  \bibinfo{author}{\bibfnamefont{D.}~\bibnamefont{Prabhakaran}},
  \bibinfo{author}{\bibfnamefont{Y.}~\bibnamefont{Joly}},
  \bibinfo{author}{\bibfnamefont{B.}~\bibnamefont{Janousova}},
  \bibinfo{author}{\bibfnamefont{C.}~\bibnamefont{Mazzoli}},
  \bibinfo{author}{\bibfnamefont{L.}~\bibnamefont{Paolasini}},
  \bibnamefont{and} \bibinfo{author}{\bibfnamefont{S.~B.}
  \bibnamefont{Wilkins}}, \bibinfo{journal}{Phys. Rev. B}
  \textbf{\bibinfo{volume}{76}}, \bibinfo{pages}{184420}
  (\bibinfo{year}{2007}).

\bibitem[{\citenamefont{Strempfer et~al.}(2008)\citenamefont{Strempfer,
  Bohnenbuck, Zegkinoglou, Aliouane, Landsgesell, Zimmermann, and
  Argyriou}}]{Strempfer}
\bibinfo{author}{\bibfnamefont{J.}~\bibnamefont{Strempfer}},
  \bibinfo{author}{\bibfnamefont{B.}~\bibnamefont{Bohnenbuck}},
  \bibinfo{author}{\bibfnamefont{I.}~\bibnamefont{Zegkinoglou}},
  \bibinfo{author}{\bibfnamefont{N.}~\bibnamefont{Aliouane}},
  \bibinfo{author}{\bibfnamefont{S.}~\bibnamefont{Landsgesell}},
  \bibinfo{author}{\bibfnamefont{M.~v.} \bibnamefont{Zimmermann}},
  \bibnamefont{and} \bibinfo{author}{\bibfnamefont{D.~N.}
  \bibnamefont{Argyriou}}, \bibinfo{journal}{Phys. Rev. B}
  \textbf{\bibinfo{volume}{78}}, \bibinfo{pages}{024429}
  (\bibinfo{year}{2008}).

\bibitem[{\citenamefont{Wilkins et~al.}(2009)\citenamefont{Wilkins, Forrest,
  Beale, Bland, Walker, Mannix, Yakhou, Prabhakaran, Boothroyd, Hill
  et~al.}}]{Wilkins09}
\bibinfo{author}{\bibfnamefont{S.~B.} \bibnamefont{Wilkins}},
  \bibinfo{author}{\bibfnamefont{T.~R.} \bibnamefont{Forrest}},
  \bibinfo{author}{\bibfnamefont{T.~A.~W.} \bibnamefont{Beale}},
  \bibinfo{author}{\bibfnamefont{S.~R.} \bibnamefont{Bland}},
  \bibinfo{author}{\bibfnamefont{H.~C.} \bibnamefont{Walker}},
  \bibinfo{author}{\bibfnamefont{D.}~\bibnamefont{Mannix}},
  \bibinfo{author}{\bibfnamefont{F.}~\bibnamefont{Yakhou}},
  \bibinfo{author}{\bibfnamefont{D.}~\bibnamefont{Prabhakaran}},
  \bibinfo{author}{\bibfnamefont{A.~T.} \bibnamefont{Boothroyd}},
  \bibinfo{author}{\bibfnamefont{J.~P.} \bibnamefont{Hill}},
  \bibnamefont{et~al.}, \bibinfo{journal}{Phys. Rev. Lett.}
  \textbf{\bibinfo{volume}{103}}, \bibinfo{pages}{207602}
  (\bibinfo{year}{2009}).

\bibitem[{\citenamefont{Fabrizi et~al.}(2009)\citenamefont{Fabrizi, Walker,
  Paolasini, de~Bergevin, Boothroyd, Prabhakaran, and McMorrow}}]{Fabrizi}
\bibinfo{author}{\bibfnamefont{F.}~\bibnamefont{Fabrizi}},
  \bibinfo{author}{\bibfnamefont{H.~C.} \bibnamefont{Walker}},
  \bibinfo{author}{\bibfnamefont{L.}~\bibnamefont{Paolasini}},
  \bibinfo{author}{\bibfnamefont{F.}~\bibnamefont{de~Bergevin}},
  \bibinfo{author}{\bibfnamefont{A.~T.} \bibnamefont{Boothroyd}},
  \bibinfo{author}{\bibfnamefont{D.}~\bibnamefont{Prabhakaran}},
  \bibnamefont{and} \bibinfo{author}{\bibfnamefont{D.~F.}
  \bibnamefont{McMorrow}}, \bibinfo{journal}{Phys. Rev. Lett.}
  \textbf{\bibinfo{volume}{102}}, \bibinfo{pages}{237205}
  (\bibinfo{year}{2009}).

\bibitem[{\citenamefont{Blume and Gibbs}(1988)}]{Blume}
\bibinfo{author}{\bibfnamefont{M.}~\bibnamefont{Blume}} \bibnamefont{and}
  \bibinfo{author}{\bibfnamefont{D.}~\bibnamefont{Gibbs}},
  \bibinfo{journal}{Phys. Rev. B} \textbf{\bibinfo{volume}{37}},
  \bibinfo{pages}{1779} (\bibinfo{year}{1988}).

\bibitem[{\citenamefont{de~Bergevin and Brunel}(1981)}]{deBergevin}
\bibinfo{author}{\bibfnamefont{F.}~\bibnamefont{de~Bergevin}} \bibnamefont{and}
  \bibinfo{author}{\bibfnamefont{M.}~\bibnamefont{Brunel}},
  \bibinfo{journal}{Acta Cryst.} \textbf{\bibinfo{volume}{A37}},
  \bibinfo{pages}{314} (\bibinfo{year}{1981}).

\bibitem[{\citenamefont{Lawes et~al.}(2004)\citenamefont{Lawes, Kenzelmann,
  Rogado, Kim, Jorge, Cava, Aharony, Entin-Wohlman, Harris, Yildirim
  et~al.}}]{Lawes04}
\bibinfo{author}{\bibfnamefont{G.}~\bibnamefont{Lawes}},
  \bibinfo{author}{\bibfnamefont{M.}~\bibnamefont{Kenzelmann}},
  \bibinfo{author}{\bibfnamefont{N.}~\bibnamefont{Rogado}},
  \bibinfo{author}{\bibfnamefont{K.~H.} \bibnamefont{Kim}},
  \bibinfo{author}{\bibfnamefont{G.~A.} \bibnamefont{Jorge}},
  \bibinfo{author}{\bibfnamefont{R.~J.} \bibnamefont{Cava}},
  \bibinfo{author}{\bibfnamefont{A.}~\bibnamefont{Aharony}},
  \bibinfo{author}{\bibfnamefont{O.}~\bibnamefont{Entin-Wohlman}},
  \bibinfo{author}{\bibfnamefont{A.~B.} \bibnamefont{Harris}},
  \bibinfo{author}{\bibfnamefont{T.}~\bibnamefont{Yildirim}},
  \bibnamefont{et~al.}, \bibinfo{journal}{Phys. Rev. Lett.}
  \textbf{\bibinfo{volume}{93}}, \bibinfo{pages}{247201}
  (\bibinfo{year}{2004}).

\bibitem[{\citenamefont{Lawes et~al.}(2005)\citenamefont{Lawes, Harris, Kimura,
  Rogado, Cava, Aharony, Entin-Wohlman, Yildirim, Kenzelmann, Broholm
  et~al.}}]{Lawes05}
\bibinfo{author}{\bibfnamefont{G.}~\bibnamefont{Lawes}},
  \bibinfo{author}{\bibfnamefont{A.~B.} \bibnamefont{Harris}},
  \bibinfo{author}{\bibfnamefont{T.}~\bibnamefont{Kimura}},
  \bibinfo{author}{\bibfnamefont{N.}~\bibnamefont{Rogado}},
  \bibinfo{author}{\bibfnamefont{R.~J.} \bibnamefont{Cava}},
  \bibinfo{author}{\bibfnamefont{A.}~\bibnamefont{Aharony}},
  \bibinfo{author}{\bibfnamefont{O.}~\bibnamefont{Entin-Wohlman}},
  \bibinfo{author}{\bibfnamefont{T.}~\bibnamefont{Yildirim}},
  \bibinfo{author}{\bibfnamefont{M.}~\bibnamefont{Kenzelmann}},
  \bibinfo{author}{\bibfnamefont{C.}~\bibnamefont{Broholm}},
  \bibnamefont{et~al.}, \bibinfo{journal}{Phys. Rev. Lett.}
  \textbf{\bibinfo{volume}{95}}, \bibinfo{pages}{087205}
  (\bibinfo{year}{2005}).

\bibitem[{\citenamefont{Cabrera et~al.}(2009)\citenamefont{Cabrera, Kenzelmann,
  Lawes, Chen, Chen, Erwin, Gentile, Leao, Lynn, Rogado et~al.}}]{Cabrera}
\bibinfo{author}{\bibfnamefont{I.}~\bibnamefont{Cabrera}},
  \bibinfo{author}{\bibfnamefont{M.}~\bibnamefont{Kenzelmann}},
  \bibinfo{author}{\bibfnamefont{G.}~\bibnamefont{Lawes}},
  \bibinfo{author}{\bibfnamefont{Y.}~\bibnamefont{Chen}},
  \bibinfo{author}{\bibfnamefont{W.~C.} \bibnamefont{Chen}},
  \bibinfo{author}{\bibfnamefont{R.}~\bibnamefont{Erwin}},
  \bibinfo{author}{\bibfnamefont{T.~R.} \bibnamefont{Gentile}},
  \bibinfo{author}{\bibfnamefont{J.~B.} \bibnamefont{Leao}},
  \bibinfo{author}{\bibfnamefont{J.~W.} \bibnamefont{Lynn}},
  \bibinfo{author}{\bibfnamefont{N.}~\bibnamefont{Rogado}},
  \bibnamefont{et~al.}, \bibinfo{journal}{Phys. Rev. Lett.}
  \textbf{\bibinfo{volume}{103}}, \bibinfo{pages}{087201}
  (\bibinfo{year}{2009}).

\bibitem[{\citenamefont{Paolasini et~al.}(2007)\citenamefont{Paolasini,
  Detlefs, Mazzoli, Wilkins, Deen, Bombardi, Kernavanois, de~Bergevin, Yakhou,
  Valade et~al.}}]{Paolasini07}
\bibinfo{author}{\bibfnamefont{L.}~\bibnamefont{Paolasini}},
  \bibinfo{author}{\bibfnamefont{C.}~\bibnamefont{Detlefs}},
  \bibinfo{author}{\bibfnamefont{C.}~\bibnamefont{Mazzoli}},
  \bibinfo{author}{\bibfnamefont{S.}~\bibnamefont{Wilkins}},
  \bibinfo{author}{\bibfnamefont{P.~P.} \bibnamefont{Deen}},
  \bibinfo{author}{\bibfnamefont{A.}~\bibnamefont{Bombardi}},
  \bibinfo{author}{\bibfnamefont{N.}~\bibnamefont{Kernavanois}},
  \bibinfo{author}{\bibfnamefont{F.}~\bibnamefont{de~Bergevin}},
  \bibinfo{author}{\bibfnamefont{F.}~\bibnamefont{Yakhou}},
  \bibinfo{author}{\bibfnamefont{J.~P.} \bibnamefont{Valade}},
  \bibnamefont{et~al.}, \bibinfo{journal}{J. Synch. Rad.}
  \textbf{\bibinfo{volume}{14}}, \bibinfo{pages}{301} (\bibinfo{year}{2007}).

\bibitem[{\citenamefont{Scagnoli et~al.}(2009)\citenamefont{Scagnoli, Mazzoli,
  Bernard, Fondacaro, Paolasini, Detlefs, Fabrizi, and de~Bergevin}}]{Scagnoli}
\bibinfo{author}{\bibfnamefont{V.}~\bibnamefont{Scagnoli}},
  \bibinfo{author}{\bibfnamefont{C.}~\bibnamefont{Mazzoli}},
  \bibinfo{author}{\bibfnamefont{P.}~\bibnamefont{Bernard}},
  \bibinfo{author}{\bibfnamefont{A.}~\bibnamefont{Fondacaro}},
  \bibinfo{author}{\bibfnamefont{L.}~\bibnamefont{Paolasini}},
  \bibinfo{author}{\bibfnamefont{C.}~\bibnamefont{Detlefs}},
  \bibinfo{author}{\bibfnamefont{F.}~\bibnamefont{Fabrizi}}, \bibnamefont{and}
  \bibinfo{author}{\bibfnamefont{F.}~\bibnamefont{de~Bergevin}},
  \bibinfo{journal}{J. Synch. Rad.} \textbf{\bibinfo{volume}{16}},
  \bibinfo{pages}{778} (\bibinfo{year}{2009}).

\bibitem[{com()}]{comment}
\bibinfo{note}{Note that in NiO the Ni orbital moment was estimated to be
  $0.32\pm0.05\,\mu_\mathrm{B}$\cite{fernandez}.}

\bibitem[{\citenamefont{Finger et~al.}(2010)\citenamefont{Finger, Senff,
  Schmalzl, Schmidt, Regnault, Becker, Bohat\`{y}, and Braden}}]{Finger}
\bibinfo{author}{\bibfnamefont{T.}~\bibnamefont{Finger}},
  \bibinfo{author}{\bibfnamefont{D.}~\bibnamefont{Senff}},
  \bibinfo{author}{\bibfnamefont{K.}~\bibnamefont{Schmalzl}},
  \bibinfo{author}{\bibfnamefont{W.}~\bibnamefont{Schmidt}},
  \bibinfo{author}{\bibfnamefont{L.~P.} \bibnamefont{Regnault}},
  \bibinfo{author}{\bibfnamefont{P.}~\bibnamefont{Becker}},
  \bibinfo{author}{\bibfnamefont{L.}~\bibnamefont{Bohat\`{y}}},
  \bibnamefont{and} \bibinfo{author}{\bibfnamefont{M.}~\bibnamefont{Braden}},
  \bibinfo{journal}{Physical Review B} \textbf{\bibinfo{volume}{81}},
  \bibinfo{pages}{054430} (\bibinfo{year}{2010}).

\bibitem[{\citenamefont{Fernandez et~al.}(1998)\citenamefont{Fernandez,
  Vettier, de~Bergevin, Giles, and Neubeck}}]{fernandez}
\bibinfo{author}{\bibfnamefont{V.}~\bibnamefont{Fernandez}},
  \bibinfo{author}{\bibfnamefont{C.}~\bibnamefont{Vettier}},
  \bibinfo{author}{\bibfnamefont{F.}~\bibnamefont{de~Bergevin}},
  \bibinfo{author}{\bibfnamefont{C.}~\bibnamefont{Giles}}, \bibnamefont{and}
  \bibinfo{author}{\bibfnamefont{W.}~\bibnamefont{Neubeck}},
  \bibinfo{journal}{Phys. Rev. B} \textbf{\bibinfo{volume}{57}},
  \bibinfo{pages}{7870} (\bibinfo{year}{1998}).

\end{thebibliography}

\end{document}